\documentclass{aastex63}

\submitjournal{ApJ}
\usepackage{multirow}

\shorttitle{Eccentric Atmospheres}
\shortauthors{Mayorga et al.}

\graphicspath{{./}{figures/}}

\newcommand{\EGP}{\texttt{EGP}}
\newcommand{\picaso}{\texttt{PICASO}}
\newcommand{\TESS}{\textit{TESS}}
\newcommand{\Spitzer}{\textit{Spitzer}}
\newcommand{\Kepler}{\textit{Kepler}}

\begin{document}

\title{Variable Irradiation on 1D Cloudless Eccentric Exoplanet Atmospheres}

\correspondingauthor{L.~C. Mayorga}
\email{laura.mayorga@jhuapl.edu}
\author[0000-0002-4321-4581]{L. C. Mayorga}
\affiliation{The Johns Hopkins University Applied Physics Laboratory 11100 Johns Hopkins Rd Laurel, MD, 20723, USA}

\author[0000-0002-3196-414X]{Tyler D. Robinson}
\affiliation{Department of Astronomy and Planetary Science, Northern Arizona University, Box 6010, Flagstaff, AZ 86011, USA}
\affiliation{Habitability, Atmospheres, and Biosignatures Laboratory, Northern Arizona University, Flagstaff, AZ 86011, USA}

\author[0000-0002-5251-2943]{Mark S. Marley}
\affiliation{NASA Ames Research Center, Mountain View, CA, USA 94035}

\author[0000-0002-2739-1465]{E. M. May}
\author[0000-0002-7352-7941]{Kevin B. Stevenson}
\affiliation{The Johns Hopkins University Applied Physics Laboratory 11100 Johns Hopkins Rd Laurel, MD, 20723, USA}

\begin{abstract}
Exoplanets on eccentric orbits experience an incident stellar flux that can be markedly larger at periastron versus apoastron. This variation in instellation can lead to dramatic changes in atmospheric structure in regions of the atmosphere where the radiative and advective heating/cooling timescales are shorter than the orbital timescale. To explore this phenomenon, we develop a sophisticated one-dimensional (vertical) time-stepping atmospheric structure code, \EGP+, capable of simulating the dynamic response of atmospheric thermal and chemical structure to time-dependent perturbations. Critically, \EGP+ can efficiently simulate multiple orbits of a planet, thereby providing new opportunities for exoplanet modeling without the need for more computationally-expensive models. We make the simplifying assumption of cloud-free atmospheres, and apply our model to HAT-P-2b, HD~17156b, and HD~80606b, which are known to be on higher-eccentricity orbits. We find that for those planets which have \Spitzer{} observations, our planet-to-star ratio predictions are roughly consistent with observations. However, we are unable to reproduce the observed peak offsets from periastron passage. Finally, we discuss promising pathways forward for adding new model complexity that would enable more detailed studies of clear and cloudy eccentric planets as well as worlds orbiting active host stars.
\end{abstract}

\keywords{Exoplanet atmospheric variability (2020), Exoplanet atmospheres (487), Exoplanet dynamics (490), Star-planet interactions (2177), Eccentricity (441)}

\section{Introduction} \label{sec:intro}
The Solar System plainly demonstrates that atmospheres change: storms form and dissipate, regional weather patterns change with season, and chemical compositions change in response to outside influences. A dramatic example are the seasonal variations in surface pressure on Mars due, in part, to its eccentric orbit \added{\citep[e.g.,][]{Trainer2019SeasonalMars}}. Atmospheric variability is likely to be the norm for exoplanets, and drivers of this variability could include non-circular orbits and/or active host stars. Most universally, time-dependent phenomenon can sculpt and drive the time evolution of exoplanet atmospheres. 

With the coming of next-generation flagship missions, 30\,m class ground-based telescopes, and dedicated exoplanet instrumentation, we must establish an understanding of how time-dependent effects manifest as subtle variations in the structure of exoplanet atmospheres. The detection of time-varying atmospheric phenomenon could constrain atmospheric models. For example observing the speed at which the atmosphere responds could constrain the radiative timescale \citep[e.g. as seen with day-night pattern dependence on radiative timescale in Hot Jupiters, ][]{Showman2013DopplerJupiters, Perez-Becker2013AtmosphericJupiters, Komacek2016}, which can further constrain the heat capacity and therefore composition of the atmosphere. Likewise the omission of time-varying physics in inference models could lead to biases in interpretations \citep[e.g. in retrievals, ][]{Feng2020}.

It is common for a planet to receive variable irradiation from its host star. Radial velocity data has shown that longer-period exoplanets are more likely to be on eccentric orbits \citep{Halbwachs2005, Pont2011}. These planets are not likely rotationally-locked, are generally cooler, and are slower to respond to external atmospheric stimuli. Exoplanets on highly eccentric orbits ($e \gtrsim 0.5$) uniquely provide direct probes of key dynamical processes that shape planetary atmospheres, such chemical transitions that change the pressure at which energy is deposited in the atmosphere. The time-variable forcing experienced by these eccentric exoplanets enable the study and observation of the timescale over which atmospheres respond radiatively, dynamically, and chemically. 

Evidence for dynamic atmospheric changes with orbital distance on exoplanets has already been observed \added{\citep{Laughlin2009, Lewis2013, Lanotte2014AB}}. The \textit{Kepler} light curve of the moderately-eccentric Kepler-434b may present evidence for cloud formation, as the total system flux increases by 32\,ppm when the planet is closer than 0.1147\,AU from the host star \citetext{Dittmann, AAS 2020}. The transition between the low flux state (i.e., when the planet is more distant than 0.1147\,AU) and the high flux state takes roughly 7.7\,hours \added{as measured from the light curve}. This timescale likely corresponds to the time it takes for clouds to dissipate in the atmosphere. Atmospheric models of the planet suggest that the upper atmosphere could be heated and cooled to change the physical state of potassium and sodium compounds \citetext{Dittmann, AAS 2020}. Longer-period eccentric planets, such as HD~20782b, may be slow to respond to the changes in incident flux \citep{Kane2016}.

Eccentric planets have typically been studied using general circulation models (GCM). These kinds of simulations can investigate dynamic changes in the thermal, chemical, and dynamic state of eccentric exoplanet atmospheres as a function of latitude, longitude, and depth \citep{Kataria2013, Lewis2014, Lewis2017}. However, these sophisticated tools can be computationally intensive. To understand complex phenomena, it is often necessary and advantageous to apply models of various levels of detail to explore the importance of incorporated physics on simulated results. One-dimensional (1D) codes, which aim to understand the mean global (or day side only) atmospheric structure of a planet, are a key element of this modeling hierarchy. 

\added{One-dimensional radiative-convective equilibrium models have a long history in Solar System atmospheric studies. By the 1980s 1D models were able to reproduce the observed thermal profiles for the giant planets \citep[e.g.,][and subsequent]{Appleby1984Radiative-convectiveSaturn,Appleby1986Radiative-convectiveNeptune,Marley1999b}. These successes hinged on the long radiative timescales for cool gas giants and their rapid winds and rotation rates which result in efficient redistribution of energy around the planet. In fact, in the thermal infrared, there is no detectable day/night difference in images of these planets. Failures of models to reproduce observed profiles point to shortcomings in our understanding, for example missing atmospheric absorbers or other energy sources.} 

Pioneering work in the study of eccentric exoplanets with 1D models was presented by \citet{Iro2010}.  These authors constructed a time-stepping radiative model for exoplanet atmospheres, applied this tool to known eccentric exoplanets, and predicted observables relevant to NASA's \textit{Spitzer} telescope. By necessity this study had to make a number simplifying assumptions, such as neglecting cloud and chemical composition changes as well as the use of a constant radiative timescale for each pressure layer over an orbit.

\added{Another class of models that has been applied to eccentric planets is energy balance models, which aim to estimate temperatures from consideration of radiative time constants and wind speeds \citep[e.g.,][]{Cowan2011}. An example application is found in \citep[][]{Lewis2011}. While such models provide insight, they account neither for variations in atmospheric heating rate with altitude arising from wavelength-dependent atmospheric opacities nor the internal heat redistribution within the atmosphere by radiation. Such physics can only be captured by a complete radiative transfer model, either a complete GCM or a 1D spherically symmetric model.}

Here we present our own time-stepping model built on the self-consistent framework of \citet{Marley1999b}\,---\,a 1D radiative-convective equilibrium model\,---\,and the time-stepping capabilities added by \citet{Robinson2014}, and we apply this new tool to eccentric exoplanets. \added{We are interested in learning how atmospheric thermal strucutre and composition varies around the orbit and in identifying particularly interesting planets which may be amenable to future observational followup. Such a 1D model as we apply here is well suited to these pursuits.} In \autoref{sec:methods} we describe model heritage and present new  developments to our model, which we call \EGP+. In \autoref{sec:models} we apply \EGP+ to several known eccentric exoplanets. Then, in \autoref{sec:compare}, we use a high-resolution radiative transfer model\,---\,\picaso{} \citep{Batalha2019}\,---\, to create orbital phase curves for our systems and compare to observations. In \autoref{sec:future} and \autoref{sec:conc} we discuss additional avenues for development and potential new science questions that \EGP+ can address and conclude.

\section{Methods} \label{sec:methods}
The radiative-convective equilibrium model, \EGP{}, is a 1D (vertical) climate model for planets and brown dwarfs. The precursor model to \EGP{} was originally designed for Titan \citep{McKay1989} and, since its development, \EGP{} has been applied in other Solar System-related studies \citep{Marley1999b}. The \EGP{} suite has been extensively applied to brown dwarfs \citep[e.g.,][]{Marley1996, Burrows1997, Robinson2014} as well as giant \citep[e.g.,][]{Marley1999,Fortney2005,Fortney2007,Cahoy2010,Demory2013,Webber2015,Mayorga2019} and terrestrial \citep{Morley2013,Morley2015} exoplanets. Key model inputs include the planetary surface gravity, the atmospheric composition \citep[often assumed to come from thermo-chemical equilibrium, ][]{Saumon2008} and metallicity, the host star spectral type and insolation, the pressure at the base of the model atmosphere (set to be very large for gaseous planets), the recirculation efficiency, the condensate sedimentation efficiency \citep{Ackerman2001}, and the internal heat flux.

\citet{Robinson2014} adapted the brown dwarf branch of the model \citep{Marley1996, Marley2002} to compute time-stepping evolution in order to explore the time-dependent evolution of deep atmospheric thermal perturbations. Assuming hydrostatic equilibrium, the time-stepping code carefully tracks the radiative and convective heat input and output for each pressure layer in the 1D atmosphere. This treatment was used to explore how the deep atmosphere communicates energy to the upper atmosphere in a case study of variability in a T6.5 brown dwarf, 2MASS~J22282889-431026 \citep{Buenzli2012a}. Typically, \added{vertical} energy transport in the upper atmosphere is radiation-dominated with a transition occurring deeper in the atmosphere to transport that is dominated by convection. \deleted{Here,} Convection in the deep atmosphere is modeled by a dynamic mixing-length treatment \citep{Vitense1953, Gierasch1968}. By introducing a thermal perturbation, the temperature-pressure (TP) profile of the atmosphere responds and, over time, transfers energy either upward or downward through the atmospheric column.

Our new modeling framework, \EGP+, builds on work from \citet{Robinson2014} by incorporating updates from the exoplanet \EGP{} branch that include the absorption of incident stellar flux and accounting for a time-dependent distance from a host star. Rather than iteratively solving for an equilibrium radiative-convective solution (as is common with \EGP{}), the time-stepping code begins with an initial model atmospheric structure, computes radiative and convective energy flux divergences in each layer of the atmosphere, and then time-steps the code appropriately (e.g., atmospheric regions which gain net energy from radiative and/or convective transport will warm). Radiative heating and cooling rates are thus self-consistently calculated throughout the atmosphere. To initialize the model, we first run a cloudless chemical equilibrium \EGP{} model of the given system at the average flux distance, $d_{\overline{F_*(t)}}$,
\begin{equation}
    d_{\overline{F_*(t)}} = a(1-e^2)^{\frac{1}{4}} \,
\end{equation}
where $a$ is the orbital semi-major axis and $e$ is the orbital eccentricity \citep{Bolmont2016}. Time-stepping in \EGP+ then proceeds from this initial atmospheric structure with the planet placed at apoastron. Planet-star distances are determined from the planetary orbital elements, are computed using a Python routine, and are passed as input to \EGP+. The planetary insolation is only updated in \EGP+ when orbital distance variations have caused more than a 0.1\% variation in the incident stellar flux. \added{This is to minimize the amount of spectra calculated, while maintaining a high update rate. We discuss changes to this update rate later in this section.} Incident stellar spectra are PHOENIX models as provided by the \texttt{PySynphot} package \citep{PySynphot}.

\begin{deluxetable}{l|ccc}
\tablecaption{System properties as used for modeling purposes (data from the NASA Exoplanet Archive). \label{tbl:props}}
\tablehead{ & \colhead{HAT-P-2b\tablenotemark{a}} & \colhead{HD~17156b\tablenotemark{b}} & \colhead{HD~80606b\tablenotemark{b}}}
\startdata
$P$ (d) & 5.6335 d & 21.216 d & 111.4367 d \\
$R_\mathrm{p}$ ($R_\mathrm{J}$) & 1.157 & 1.087 & 1.003 \\
$M_\mathrm{p}$ ($M_\mathrm{J}$) & 9.09 & 3.235 & 4.116 \\
\hline
$T_\mathrm{eff}$ (K) & 6290.0 & 6079.0 & 5574.0 \\
$R_*$ ($R_\odot$) & 1.64 & 1.5 & 1.037 \\
$\log g$ & 4.16 & 4.2 & 4.4 \\
$[Fe/H]$ & 0.14 & 0.24 & 0.34 \\
\hline
$a$ (AU) & 0.06878 AU & 0.163 AU & 0.4565 AU \\
$e$ & 0.517 & 0.67 & 0.932 \\
$\omega$ ($\deg$) & 185.22 & 121.32 & 301.03 \\
\hline
age (Gyr) & 2.6 & 3.3 & 6 \\
$T_\mathrm{int}$ (K) & 340 & 202 & 190 \\
\enddata
\tablenotetext{a}{Parameters from \citet{Pal2010a}.}
\tablenotetext{b}{Parameters from \citet{Bonomo2017} with $\log g$ sourced from \citet{Southworth2011}.}
\end{deluxetable}

\begin{figure*}
    \includegraphics[width=\linewidth]{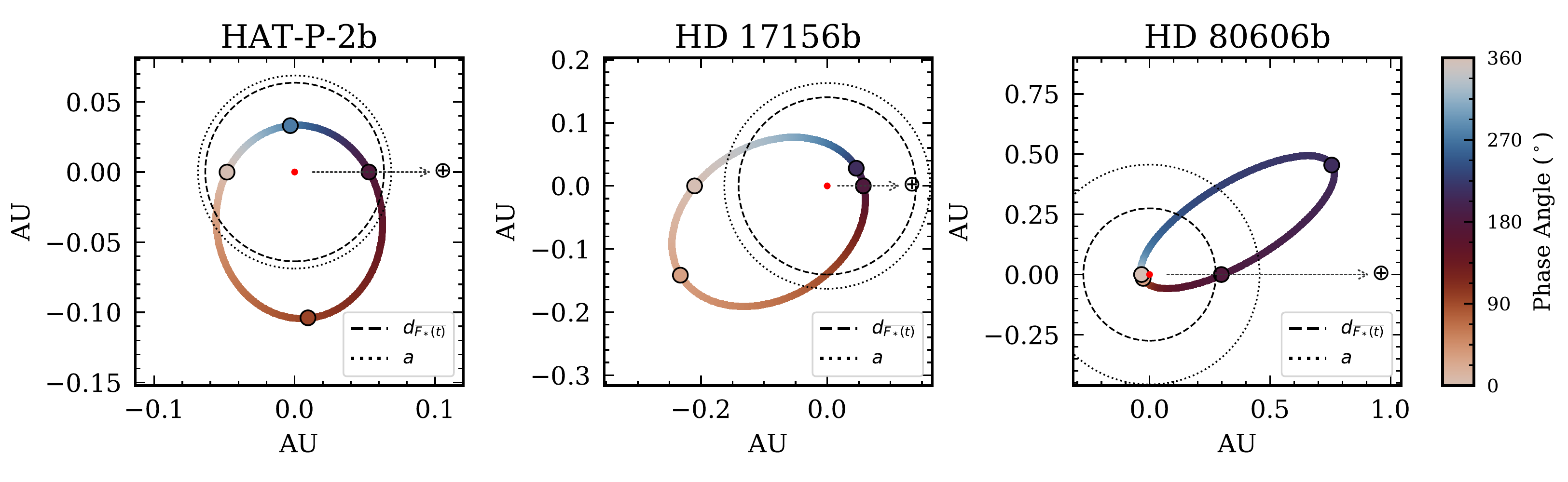}
    \caption{The orbits of our adopted eccentric planets with directionality to Earth indicated. The points are color coded by the illumination phase angle. Key orbit points (periastron, apoastron, and primary and secondary transit) are highlighted with black circles. Dashed and dotted circles mark the distances associated with the average flux, where the initial profile was calculated, and the semi-major axis of the system, respectively.}
    \label{fig:orbits}
\end{figure*}

\autoref{tbl:props} tabulates the properties as modeled for our three test cases. For this proof-of-concept demonstration, we selected planets HAT-P-2b, HD~17156b, and HD~80606b, all of which were included in the \citet{Iro2010} study. For HAT-P-2b, both \Spitzer{} light curves and GCM results have been published for comparison \citet{Lewis2013, Lewis2014}. HD~17156b results are more limited \citep{Lewis2011} as a secondary eclipse has not been observed. HD~80606b has also been widely observed and modeled \citep{deWit2016, Lewis2017}. \autoref{fig:orbits} demonstrates the planetary orbits and identifies key locations. 

All planets in this study were modeled assuming solar metallicity and C/O ratio \citetext{see Marley et al. 2021 for details}. While we assume rainout chemical equilibrium \deleted{but,} for this pilot study, we do not cold trap refractory species, such as TiO, at depth thus permitting hot stratospheres to form \citep[][]{Fortney2008} if the temperature profile crosses the condensation curve higher in the atmosphere.

We determined the appropriate internal heat flux, $\sigma T_{\rm int}^4$, from the evolution models of \citet{Marley2018} based on the mass of the companion and the rough age of the star. Simulated time steps were 5 seconds long and the vertical extent of models spanned 300\,bar to 1\,$\mu$bar. Simulated time steps are adaptively split to ensure that stability is maintained in the convective portion of the atmosphere following the Courant-Friedrichs-Lewy condition. Here, simulated time sub-steps, where convective heating and cooling rates are applied and updated, have their length adjusted so that convective motions cannot pass through any model layer within a single sub-step. 

Each model is run for 10 orbits to allow the model to settle into a quasi-steady state, the equivalent to ``spin up'' in GCMs. We determine the requisite spin up time by monitoring the percent difference in temperature at a given pressure as compared to the final orbit. In general, we see an overall warming in the atmosphere at all pressures until the quasi-steady state is reached, with the upper atmosphere adapting faster than at deeper pressures, which have greater thermal inertia. After the quasi-steady state is reached, the percent variation between orbits is only significant during periastron passages, but we find that this is correlated to our decision of when to update the stellar flux/distance to the host star. To test this, we also ran each model with a 1\% flux/distance criterion and the variation in periastron temperatures between orbits increased while our choice of 0.1\% flux/distance criterion adequately minimized this orbit-to-orbit differences. In \autoref{sec:models} we describe how HD~17156b takes the most orbits to spin up the deep atmosphere to within the range of the periastron envelope. The overall warming of all models suggests that the average flux distance may be too cold of a start for eccentric planets, while a warmer initial profile may simply allow the atmosphere to settle into the quasi-steady state sooner. In the last three orbits, which we take to be in quasi-steady state, the deep atmosphere is heating by no more than 1\% in $\Delta T/T$ in a trend that is plateauing in comparison to the earlier orbits.

For the eccentric, long-period planets \deleted{in the context of direct imaging}, tidal heating \added{for the planets considered here} will generally be negligible \added{compared to their intrinsic internal luminosity}, as tidal heating falls off with semi-major axis $a$ as $a^{-15/2}$ \citep[e.g.][]{Bodenheimer2001}. Hence\deleted{--for the planets considered here--}tidal heating is a \deleted{tiny} perturbation to their internal heat content. Furthermore, because tidal energy is deposited throughout the planetary interior \added{with a profile that varies with time and stellar distance} and this energy must reach the atmosphere by convection, the phase lag between any increment of internal heat \added{added by tides at some depth} to reach the visible atmosphere will be a \deleted{complicated function with depth} \added{complex}. For this proof of concept study, we omit tidal heating, but, given a sufficiently sophisticated model, such internal heating can be included in the same way that the flux perturbations were implemented in \citet{Robinson2014}.

Finally, we reiterate that our models are 1D and, thus, intended to represent a global average. Our simulations do not include any three-dimensional (3D) physics, such as winds, but model heat redistribution through a parameterized recirculation efficiency that approximates rotation. By computing a mean temperature profile, we are assuming that the rotation rate is faster than the time between appreciable incident flux differences. Recent studies have begun to explore ways to include other parameterized circulation effects in 1D simulations \citep{Gandhi2020}. 

\section{Model Results} \label{sec:models}

\begin{figure*}
    \includegraphics[width=\linewidth]{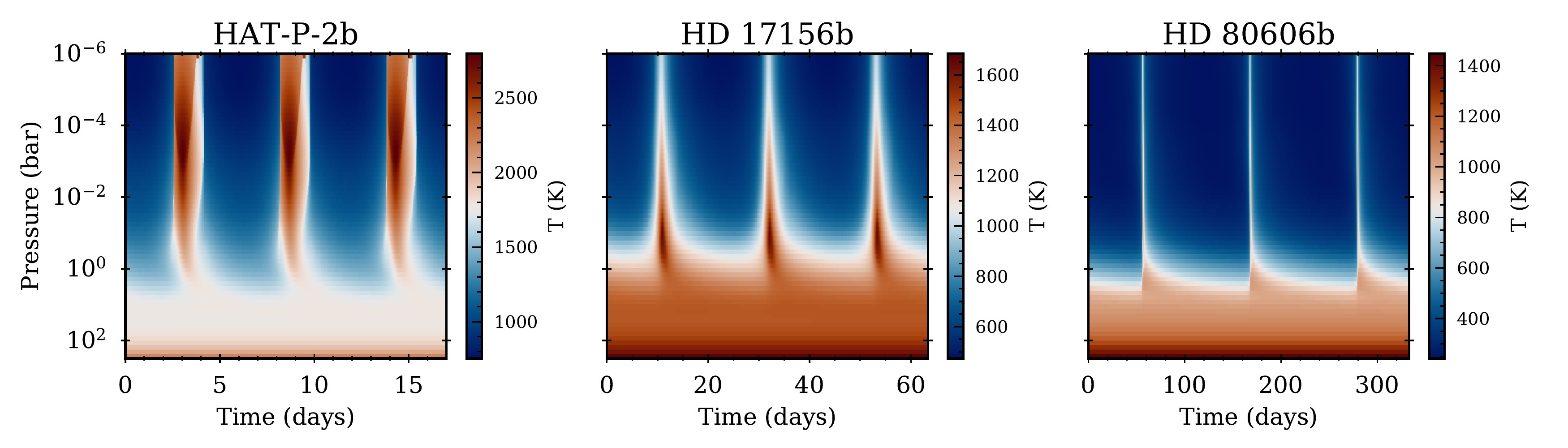}
    \caption{Temperature-pressure profiles as a function of time from apoastron for the last three simulated orbits. Hotter temperatures are represented in red and cooler temperatures in blue, as is indicated by each planet's respective (and distinct) colorbar.}
    \label{fig:2dtp}
\end{figure*}

\autoref{fig:2dtp} shows TP profiles for each planet as a function of time for the last three simulated orbits. Time sampling is variable to minimize file size, is determined by the given planet's distance to the star, and typically spans minutes to days. The deep adiabat is essentially fixed throughout the run and the upper atmosphere heats and cools in response to every periastron passage, which can be seen as the red (hotter) portions of the diagrams. 

\begin{figure*}[!bh]
    \includegraphics[width=\linewidth]{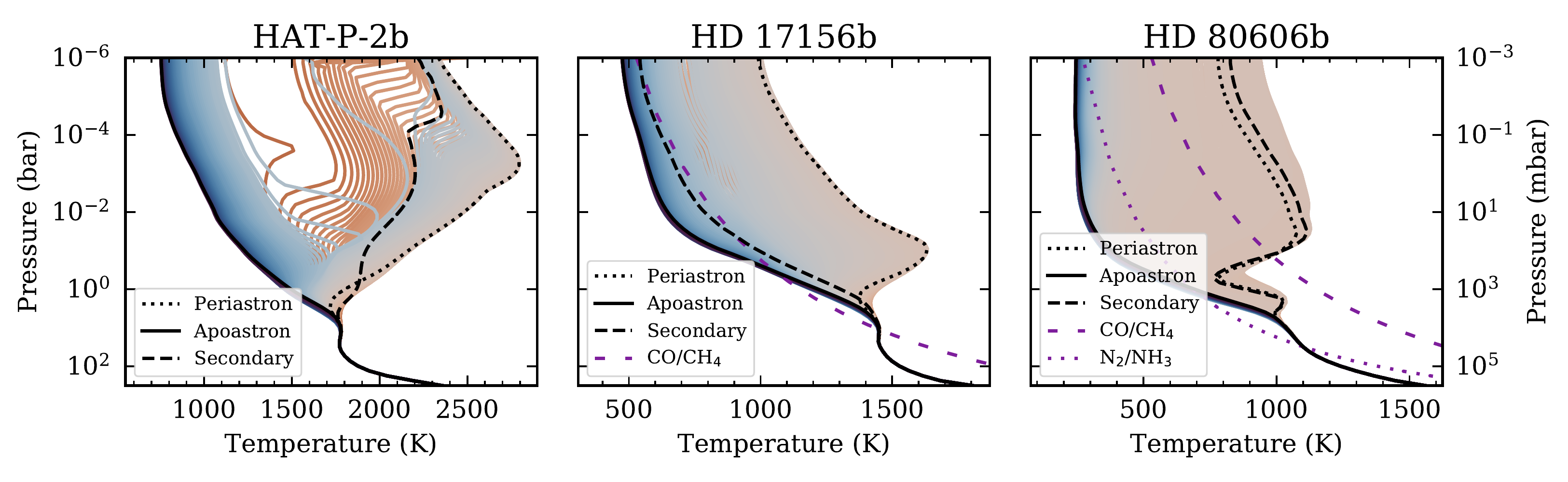}
    \caption{Temperature-pressure profiles over the course of the last complete simulated orbit. Blue colors are associated with traveling inwards from apoastron to periastron, and red colors with returning to apoastron from periastron. Dotted lines indicate the profile associated with periastron, a solid line marks the profile associated with apoastron, and the dashed line indicates where secondary eclipse occurs (as viewed from Earth). When applicable, we show the CO/CH$_4$ and N$_2$/NH$_3$ equal abundance boundaries in purple dashed and dotted lines, respectively.}
    \label{fig:tp}
\end{figure*}

Since HAT-P-2b is the closest of the three to its star and has a moderate eccentricity, its periastron passages dramatically heat the upper atmosphere, particularly at the lowest pressures. The other two planets (HD~17156b and HD~80606b) have less dramatic increases at the top of the atmosphere and, instead, their thermal profiles show greater response at pressures near 0.1--1\,bar. \autoref{fig:tp} highlights only the TP profiles from the last simulated orbit, color coded by time. Here it is easier to see the dramatic variations in temperature above roughly 1~bar. We note that the coolest profiles for each planet are nearly equivalent with the apoastron profile. Conversely, there is a delay in atmospheric heating such that the hottest profile occurs slightly after periastron passage. Our simulations show that on approach to periastron all of the models indicate the response to periastron passage is rapid, with the heating (and, thus, brightening) on approach having a sharp rise with a comparatively slower dimming on post-periastron passage.

Since these simulations are cloudless, we might expect the formation and evaporation of clouds would have additional consequences on the response time, and the magnitude of the atmospheric response, to periastron passage. For example, HD~80606b's periastron distance is nearly as close as HAT-P-2b's periastron distance, but HD~80606b remains cool and reaches only about 1200\,K. The dramatic response and temperature inversion in HAT-P-2b's atmosphere could potentially be due to TiO and VO, which are included as sources of opacity \citep{Hubeny2003,Fortney2008}.

The pressure level at which the maximum temperature response is seen is inversely correlated with each planet's gravity. HD~17156b has the lowest gravity and HAT-P-2b has the highest gravity. HAT-P-2b TP profiles show the most variation near 1\,mbar, while HD~80606b is closer to 10\,mbar and HD~17156b is closer to 100\,mbar. One would expect \deleted{from different gravities} that the higher gravity, i.e. less dense, atmospheres respond at greater pressure depths in the atmosphere. This indicates that the dominant factor in determining the magnitude and observability of a thermal response in an atmosphere is not the distance from the star at periastron or the planet's gravity, but instead some other variable or combination of variables, such as the timescale of periastron passage vs the radiative timescale of the upper atmosphere. The radiative timescale \added{of an atmospheric layer of thickness $\Delta p$}, $\tau_\mathrm{rad}$, is given by
\begin{equation}
\tau_\mathrm{rad}=\frac{\Delta p}{g}\frac{c_p}{4 \sigma T^3}
\end{equation}
where \added{$\Delta$}$p$ is the pressure \added{difference between the top and bottom of the layer}, $T$ is the temperature, $g$ is the surface gravity, and $c_p$ is the heat capacity. The radiative timescale in HAT-P-2b's atmosphere is the shortest of the three.

\begin{figure*}
    \includegraphics[width=\linewidth]{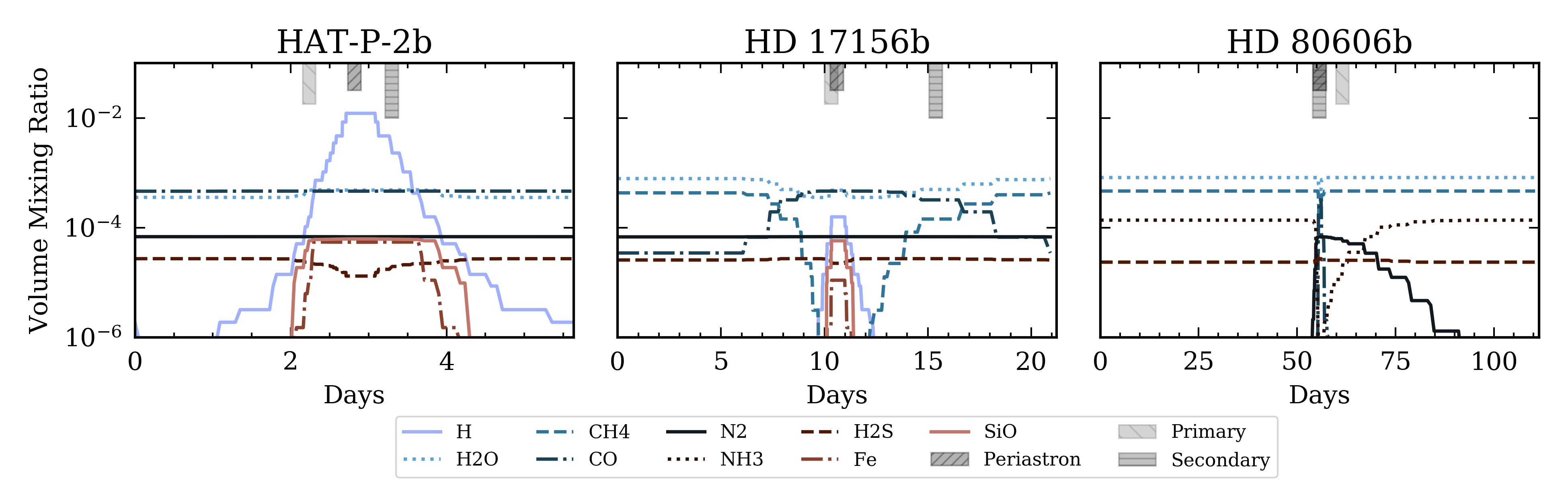}
    \caption{The volume mixing ratios at 100\,mbar over the course of the last complete orbit modeled for the dominant species in the atmosphere. Vertical hatched regions mark the times of key orbit positions, such as periastron ('//'), secondary transit ('-'), and primary transit ('\textbackslash\textbackslash'). Note that we have excluded H$_2$ and He, which are constant throughout the orbit.}
    \label{fig:mol}
\end{figure*}

We use \picaso{} \deleted{\citep{Batalha2019}} to generate spectra and light curves for any given planet at any point in its orbit \added{and phase angle}. While HAT-P-2b is quite hot, the other two planets are cool enough such that we can expect them to cross the CO/CH$_4$ and N$_2$/NH$_3$ equal-abundance boundaries \citep{Visscher2012, Fortney2020}. These chemical changes affect the atmosphere's ability to cool rapidly after periastron passage. \autoref{fig:mol} shows the volume mixing ratios of the dominant species \added{at 100\,mbar} in our modeled atmospheres over the course of the last simulated orbit. \added{For example, in HAT-P-2b is warm enough that atomic hydrogen becomes the third most important species, after H$_2$ and He, near periastron.} Note that \replaced{these results}{this figure} omits H$_2$ and He, which have constant mixing ratios for all three models. In particular, we can see that HAT-P-2b is dominated by CO throughout its entire orbit, while both HD~17156b and HD~80606b are dominated by CH$_4$ near apoastron which is then converted to CO as the planets warm. In fact, our model of HD~80606b is only dominated by CO for a few days after periastron passage,  showing a quick rise in CO and a comparatively slower return to CH$_4$-dominated conditions. In HD~80606b, the transition from NH$_3$ to N$_2$ is longer-lived. We stress that the analysis here assumes complete chemical equilibrium. If the chemical equilibrium timescale \citep[e.g.][]{Zahnle2016} is longer than the orbital thermal evolution timescale, then chemical equilibrium will not hold.

\begin{figure*}
    \includegraphics[width=\linewidth]{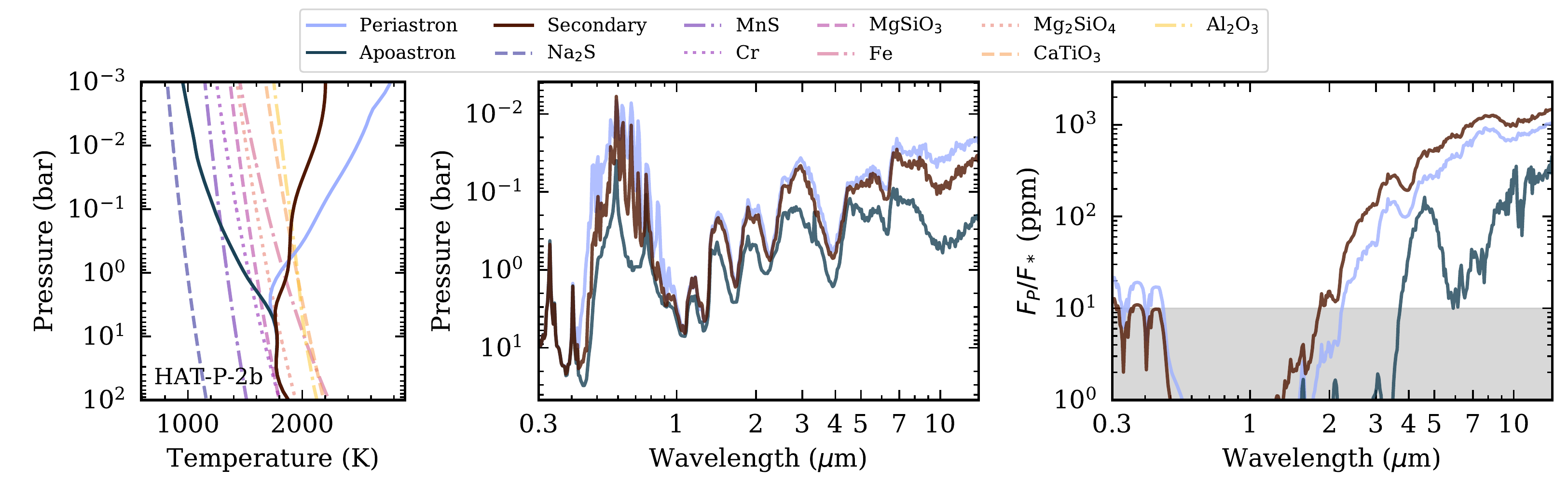}
    \includegraphics[width=\linewidth]{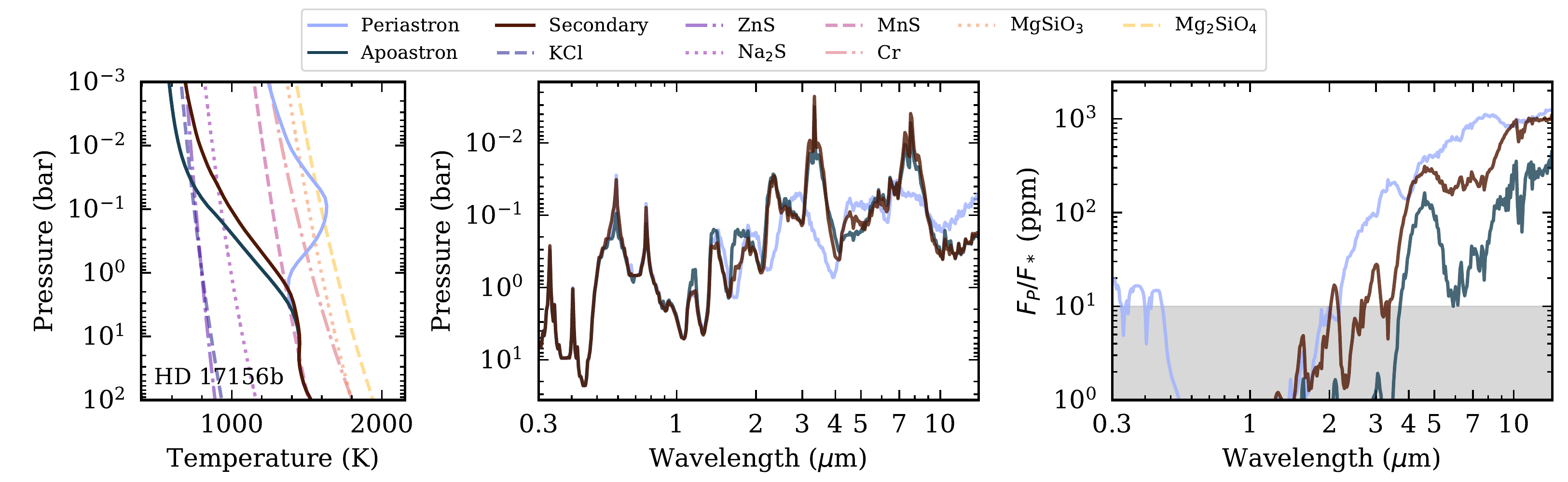}
    \includegraphics[width=\linewidth]{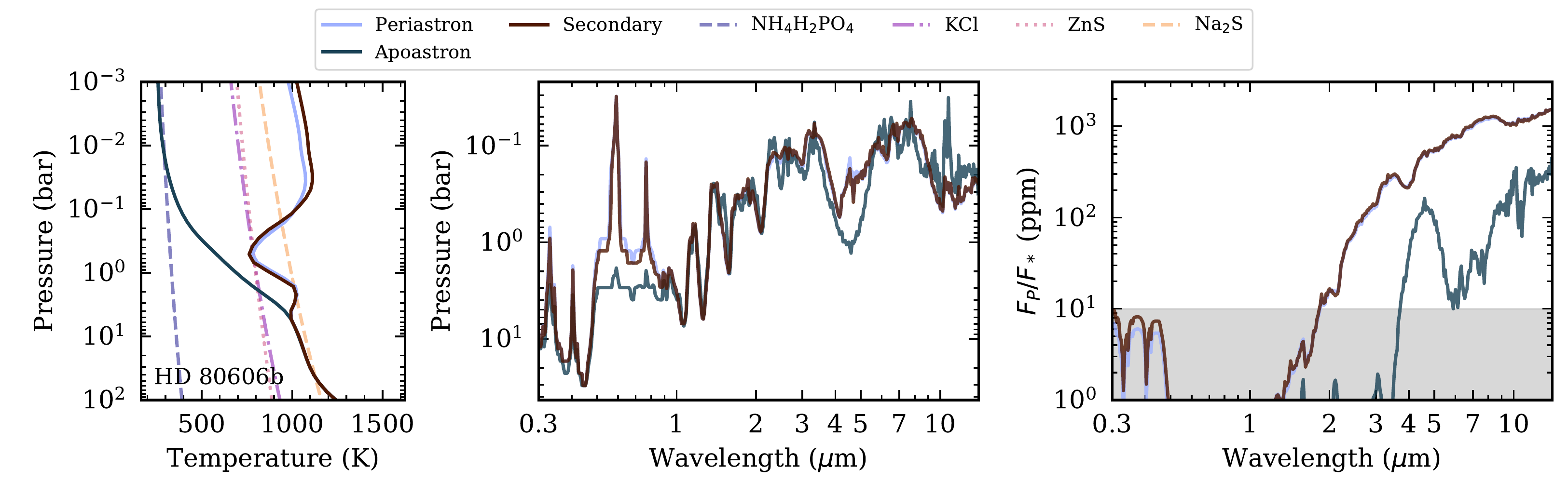}
    \caption{Temperature-pressure profiles, pressure of the $\tau = 0.5$ layer, and planet-star flux ratio spectra at key orbital locations for simulated planets. Each planet occupies a different row (HAT-P-2b on top, HD~17156b in the middle, and HD~80606b on bottom). Note that the legends are unique to each row. Left column: the TP profiles for the planet at periastron, apoastron, and at full phase (secondary eclipse) overplotted on cloud condensation curves that could potentially form. Middle column: photon attenuation diagram showing the pressure at which the optical depth reaches 0.5, indicating the atmospheric location where spectra have greatest sensitivity. The Rayleigh slope limiting the pressure being probed at short wavelengths is omitted \added{from the figure} to highlight chemical changes in the atmosphere. Right column: the planet-star flux ratio as a function of wavelength for each orbit position. The grayed out region marks a 10\,ppm precision limit for present day instrumentation where spectral signatures could likely not be observed for each planet.}
    \label{fig:planets}
\end{figure*}

\autoref{fig:planets} summarizes observable quantities for each planet at key orbital positions: periastron, apoastron, and when the planet is at full phase (nominally the secondary eclipse). The pressure being probed in the atmosphere of HD~80606b is fairly similar no matter when it is being observed, even though the radiative timescale at those pressures is changing throughout the orbit. The radiative timescale should be observed to be longest at short wavelengths for all planets at $\lambda<$ 0.5\,\micron{}, but can range from seconds to hundreds of hours from 0.3--14\,\micron{} depending on orbital position. Our models were run without clouds, but we can see in the left panels of \autoref{fig:planets} that we might expect a variety of clouds at all orbital positions. The addition of clouds could significantly change the pressure level that observation would probe in the atmosphere and sequester species deeper in the atmosphere so that they are no longer observable. Clouds such as CaTiO$_3$ in HAT-P-2b could remove TiO from the upper atmosphere, silicate clouds in HD~17156b could also cold trap these silicates, and Na$_2$S clouds could affect the atmospheric structure of HD~80606b.

For HAT-P-2b and HD~17156b the simulations show a greater variation in the pressure level being probed at these orbit points than HD~80606b. HAT-P-2b and HD~17156b also show the greatest variation in the planet-star flux ratio, $F_\mathrm{P}/F_*$, over these orbit positions as a function of wavelength, which is shown in the right panels of \autoref{fig:planets}. This includes both reflected and thermal emission. \deleted{All the models show that peak brightness occurs after periastron. So for planets where the secondary eclipse occurs after, but near, periastron passage, we expect these worlds to be brighter than at other perspectives. This is particularly true for HD~80606b, where the peak flux occurs some 15--30\,hrs after periastron, which is closely aligned with the secondary eclipse position (see sec:compare). Therefore the full dynamic range in HD~80606b is not  captured in this set of figures.}

At short wavelengths, our models indicate that observing flux from these planets may be challenging as the signals we predict (\textless 10\,ppm) are small. Our predictions are that, instead, longer wavelength observations would present better observing conditions and may be able to identify spectral variations due to dynamic chemistry near 6--7\,\micron{} such as those that can be done by the MIRI instrument on \emph{JWST}.

\section{Comparison with Observations} \label{sec:compare}
We can compare the results of our modeling efforts with other modeling studies and prior observations of these planets. Typically, the temperatures predicted by \citet{Iro2010} are much cooler than those predicted here and than those suggested by observations \added{because their models are purely radiative}. We expect some discrepancy when compared against actual data due to the absence of clouds in our simulated atmospheres, which were also not included in \citet{Iro2010}.

\begin{figure*}
    \includegraphics[width=\linewidth]{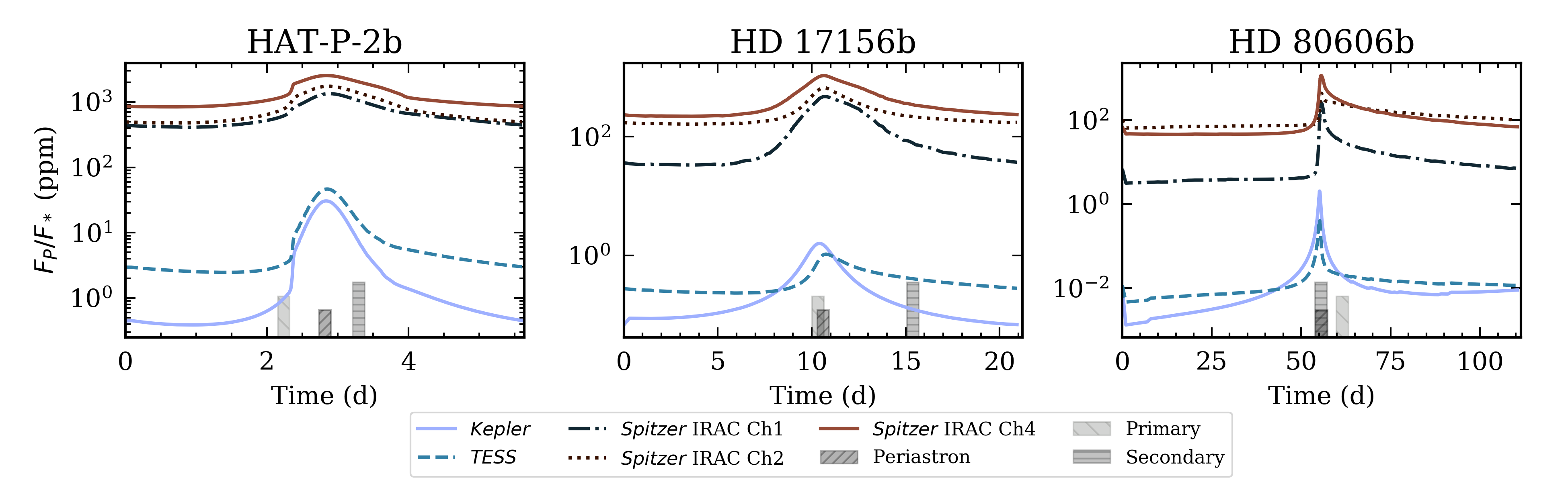}
    \caption{The flux ratio expected in selected bandpasses. The different bandpasses are colored from blue to red depending on their effective wavelengths. Vertical hatched regions mark the times of key orbit positions, such as periastron ('//'), secondary transit ('-'), and primary transit ('\textbackslash\textbackslash').}
    \label{fig:light}
\end{figure*}

These planets have prior observations in a variety of bandpasses and may be observed with future instrumentation as well. To compare against these observations we filter-integrate our flux ratios to predict the observed flux ratio in the given bandpass. We have selected the following bandpasses for comparison: the \Kepler{} bandpass, the \emph{Transiting Exoplanets Survey Satellite} (\TESS{}) bandpass, and \Spitzer{} channels 1, 2, and 4 (3.6\,\micron{}, 4.5\,\micron{}, and 8.0\,\micron). We show in \autoref{fig:light} the light curves as predicted by our simulations. In \autoref{tbl:data} we tabulate the \replaced{peak-to-peak}{amplitude of the} variation in the flux ratio and the peak offset delay after periastron in days, \added{and summarize the comparison observations. Reflecting the complex atmospheric response some bandpasses peak before periastron, which is denoted as a negative entry in the table}.

\begin{deluxetable*}{l|l|rr||cc|r}
\tablecaption{The predicted \replaced{peak-to-peak}{amplitude of the} planet-star flux ratio and peak offset from periastron by bandpass for each planet as compared to model or measured predictions from prior works. \label{tbl:data}}
\tablehead{\colhead{Planet} & \colhead{Filter} & \multicolumn{2}{c}{Predictions (This Work)} & \multicolumn{2}{c}{Observations (Prior Works)} & \colhead{Source}\\
\cline{3-6}  &  & $F_\mathrm{P}/F_*$ (ppm) & Offset (hrs) & $F_\mathrm{P}/F_*$ (ppm) & Offset (hrs) & }
\startdata
\multirow{5}{*}{HAT-P-2b}  &  \Kepler{}   &  30  &  0.6 & & & \\
  &  \TESS{}   &  44  &  1.0 & & & \\
  &  \Spitzer{} IRAC Ch1  &  930  &  1.7 & 1138$\pm$89 & 4.39$\pm$0.28 & \citet{Lewis2013} \\
  &  \Spitzer{} IRAC Ch2  &  1263  &  1.3 & 1162$\pm$80 & 5.84$\pm$0.39 & \citet{Lewis2013} \\
  &  \Spitzer{} IRAC Ch4  &  1695  &  0.5 & 1888$\pm$72 & 4.68$\pm$0.37 & \citet{Lewis2013} \\
\hline
\multirow{5}{*}{HD17156b}  &  \Kepler{}   &  2  &  -4.7 & & & \\
  &  \TESS{}   &  1  &  3.3 & & & \\
  &  \Spitzer{} IRAC Ch1  &  443  &  2.1 & & & \\
  &  \Spitzer{} IRAC Ch2  &  506  &  1.1 & & & (prediction) \\
  &  \Spitzer{} IRAC Ch4  &  851  &  0.8 & $\sim$500 & $\sim$30 & \citet{Irwin2008}\\
\hline
\multirow{5}{*}{HD80606b}  &  \Kepler{}   &  2  &  -14.2 & & & \\
  &  \TESS{}   &  0  &  -14.2 & & & \\
  &  \Spitzer{} IRAC Ch1  &  257  &  -1.6 & & & \\
  &  \Spitzer{} IRAC Ch2  &  375  &  -2.2 & 738$\pm$52 & $\sim$ 0 & \citet{deWit2016}\\
  &  \Spitzer{} IRAC Ch4  &  1118  &  -2.7 & 1105$\pm$121 & $\sim$ 0 & \citet{deWit2016} \\
\enddata
\end{deluxetable*}

From the \citet{Lewis2013} \Spitzer{} data, the phase curve of HAT-P-2b was measured to have a maximum flux offset relative to the time of periastron passage of 4.39\,hr in the 3.6\,\micron{} channel and 5.84\,hr in the 4.5\,\micron{} channel. Our predicted offsets of roughly 1.7\,hr for channel 1 and 1.3\,hr in the 4.5\,\micron{} channel are much smaller. Further analysis by \citet{Lewis2014} concluded that the exact timing of the peak flux could not be explained with a single GCM model and suggested that additional physics were required, such as disequilibrium chemistry, pressure-dependent drag effects, or non-solar atmospheric abundances. Despite not including the day/night effects captured in GCM models, our model predictions are within 2--3$\sigma$ of the measured maximum fluxes in channels 1, 2, and 4.

Hydrodynamical simulations, in 2D, of HAT-P-2b were conducted by \citet{Langton2008} and also predicted fluxes in \Spitzer{} channel 4. From that work the maximum flux was given to be 900\,ppm, which is much less than that observed by \Spitzer{} and subsequent modeling, and an offset of roughly 9--10\,hrs from periastron, which is larger than that predicted here and measured by \citet{Lewis2013}.

In \citet{Lewis2011}, 3D GCM results of HD~17156b predicted that at 30\,mbar there would be a delay of 3\,days from periastron for the planet to be at its hottest, but did not specify peak fluxes. In none of our bandpasses do we see an offset greater than 4.7\,hrs. Not only is the secondary eclipse for this planet unlikely to be observed \citep{Gillon2008}, but the flux variation in the optical is also currently out of reach of present instrumentation although in the far infrared this may become more accessible in the future. \citet{Irwin2008} used the same 2D hydrodynamical model of \citet{Langton2008} and applied it to HD~17156b, which had not been included in the original work, to predict that the \Spitzer{} channel 4 peak flux would be approximately 500\,ppm approximately 30\,hrs after periastron passage. Our predictions are for a much brighter peak flux of 851\,ppm. An attempt was made to observe the 8\,\micron{} flux of HD~17156b, but no results were published \citetext{PI: Croll, Proposal ID: 50747}. This is likely due to strong ramps seen in 8\,\micron{} \Spitzer{} data that lead to difficulties in calibrating these multi-visit observations \citep{Agol2010}. \emph{JWST} may finally settle the debate about the cooling and transport of energy on this planet.

\citet{Langton2008} also applied their 2D hydrodynamical model to HD~80606b and predicted a \Spitzer{} channel 4 maximum flux of 800\,ppm, which is smaller than that predicted here and measured by \citet{deWit2016}. \Spitzer{} light curves of HD~80606b were published by \citet{deWit2016} for the 4.5\,\micron{} and 8.0\,\micron{} channels. The \replaced{peak-to-peak}{peak} flux ratio in the 4.5\,\micron{} band was nearly 750\,ppm, double our model predictions. This discrepancy can likely be explained by inhomogeneous cloud coverage \citep{Lewis2017}. \citet{Langton2008} also predict that the peak may be ever so slightly after periastron passage, with another resurgence in flux almost 40\,hrs later. The \Spitzer{} observations also suggest that the peak in the light curve should occur at periastron if not even a few hours before, but this is not addressed in \citet{deWit2016}. Simulations from \citet{Lewis2017} determined that the peak varies with the assumed planetary rotation period. In \Spitzer{} channel 4 the peak flux occurs 0.5\,hrs before periastron passage to 1.3\,hrs after (-0.5--1.3\,hrs) periastron passage depending on the rotation period assumed. Our prediction is even earlier for this channel at 2.7 hours before. In channel 2, \citet{Lewis2017} predict the peak flux occurs at a range from 0.2--1.6\,hrs after periastron and we predict -2.2\,hrs.

\section{Future Developments} \label{sec:future}
The spectral properties of, for example, a close-in tidally-locked planet's reflected light depend on the varying atmospheric chemistry driven by the diurnal variations in the atmosphere. Time-resolved photometry/spectroscopy is the only way to assess properties of non-homogeneous clouds and the distribution of such clouds has already been shown to cause degeneracies in the modeling of transmission spectra \citep{ApaiSAG15, Line2016}. 

The \EGP{} framework allows for the addition of radiatively-active and self-consistent clouds. It is also being continuously developed to include additional physical complexity, such as disequilibrium chemistry and non-H/He-dominated atmospheres, which were ignored and excluded here. Updates in chemistry or opacities can be continuously incorporated as they become available. There is some uncertainty in our simulations for when the peak actually occurs due to the way we update the distance/stellar flux. Updating the stellar flux more frequently would remove these inherent uncertainties as well as reduce inter-orbit variations caused by the step function at the expense of run times.

\EGP+ has a number of speed advantages over traditional 3D GCMs. We fully expect that run times will increase with the addition of clouds. Our clear atmosphere run for HD~80606b, which was for more than 1100\,days (10 orbits), took less than 11\,hrs, while HAT-P-2b took just under two hours to complete the requested 56\,days. On our CPU (Intel Core i9 9900K), the starting \EGP{} model can take about 15\,min for clear atmospheres and 30\,min--1\,hr for cloudy atmospheres depending on the profile initialized or how difficult it can be to achieve convergence with the cloud properties in the area near the convection zone and radiative zone boundary. We suspect that a similar run time increase will also occur in \EGP+ when clouds are turned on as the cloud is also adjusted with the changing atmosphere. Optimizations still remain to be implemented in either \EGP{} or \EGP+, for example, neither is parallelized. These speed advantages would allow us to quickly explore a range of model parameters, such as higher metallicities, other C/O ratios, and test how variable instellation affects planets of different gravities on eccentric orbits and even around variable stars.

\section{Conclusions} \label{sec:conc}
We report the development on a new time stepping one-dimensional atmosphere model for planets on eccentric orbits. This model bridges a gap between GCM calculations, which can have long run times, and traditional radiative-convective equilibrium atmosphere models. This new expansion in the \citet{Marley1999b} framework, which we call \EGP+, is an evolution on the dynamic 1D \citet{Robinson2014} brown dwarf modeling, but applied to extrasolar giant planets in eccentric orbits around their host star. We presented simulations for planets HAT-P-2b, HD~17156b, and HD~80606b and simulated observables across a broad wavelength range. These models were run for 10 orbits and we highlight the results from the final three orbits, after the atmosphere has reached a quasi-steady state.

In the simulations of these three planets, we see that the location of maximum temperature variation in the profile is inversely correlated with the gravity of the planet, higher gravity means the response is higher in the atmosphere at lower pressures, somewhat contrary to expectations from column mass density and opacity alone. Thus, under conditions of varying incident radiation, the lower \replaced{density}{opacity} atmospheres of higher gravity planets show greater dynamic range in their temperature responses.

We find that the simulations produce star-to-planet flux ratios that are consistent with the measurements made by \Spitzer{}, particularly in channel 4. Due to our current stellar flux computation scheme \added{and our inability to capture diurnal variations}, we are unable to reproduce the appropriate offset of the peak from periastron. More focused simulations (zoom-ins) on just the time period associated with periastron passage may be able to better reproduce the temporal offsets found by prior works. While the approach shown here may not perfectly capture all the physics taking place near periastron at the hottest temperatures and shortest radiative timescale, it has great potential for examining other effects at more distant portions of the orbit and exploring chemical timescales and the onset of chemical changes like the transition in the dominant carbon or nitrogen bearing molecule. \added{Certainly, capturing detailed variations near periastron would require additional complexity, such as advection and 2D winds.}

Such 1D time-stepping models are an important tool to rapidly explore the large parameter space necessary to determine the driving factors that control how an atmosphere responds to variable external radiation. \added{It is necessary to understand }the impact of the atmospheric response on observations spanning transit and secondary eclipse spectra from \emph{JWST}, high-resolution cross-correlation spectroscopy by ground-based instruments like G-clef \citep{Szentgyorgyi2016} and CRIRES+ \citep{Follert2014}, and direct imaging reflectance spectroscopy from missions like the \emph{Large UV, Optical, and InfraRed Explorer} (\emph{LUVOIR}) \citep{TheLUVOIRTeam2019}, and the \emph{Habitable Exoplanet Observatory} (\emph{HabEx}) \citep{HabEx}.

\acknowledgements
L.C.M, E.M.M, and K.B.S acknowledge support from APL's Independent Research and Development Program. We also thank N. Lewis for useful discussions in comparing 1D and 3D simulations. T.D.R gratefully acknowledges support from NASA's Exoplanets Research Program (No.~80NSSC18K0349). This work could not have been completed without the fundamental opacity work by Richard Freedman and Roxana Lupu. This research has made use of the NASA Exoplanet Archive, which is operated by the California Institute of Technology, under contract with the National Aeronautics and Space Administration under the Exoplanet Exploration Program.

\software{ \picaso{} \citep{Batalha2019}, Astropy \citep{Astropy}, PySynphot \citep{PySynphot}, NumPy \citep{NumPy}, Pandas \citep{Pandas}, Matplotlib \citep{matplotlib}}

\facilities{NASA Exoplanet Archive}

\bibliography{references}{}
\bibliographystyle{aasjournal}

\end{document}